\documentclass[pra,%
 aps,twocolumn,%
 amsmath,amssymb,
 superscriptaddress,
 reprint,%
]{revtex4-1}

\usepackage{dsfont}
\usepackage[utf8]{inputenc}
\usepackage{float}
\usepackage{placeins}
\usepackage{qcircuit}
\usepackage{braket}
\usepackage{amsmath}
\usepackage[normalem]{ulem}
\usepackage{todonotes}
\usepackage{xcolor}
\usepackage{multirow}
\usepackage{braket}
\usepackage{bm}
\usepackage{bbold}
\usepackage{todonotes}
\usepackage{wrapfig}
\usepackage{appendix}
\usepackage{graphicx,lipsum}

\makeatletter
\renewcommand{\widetext@grid}{}
\makeatother

\usepackage{multirow}
\usepackage{makecell} 
\usepackage{tabularray}

\usepackage{graphicx}
\usepackage{dcolumn}
\usepackage{bm}

\usepackage{etoolbox}
\usepackage{braket}
\usepackage{hyperref}
\usepackage{algorithm}
\usepackage{algpseudocode}
\usepackage{setspace}
\usepackage{float}
\usepackage{placeins}
\usepackage{multirow}
\usepackage{booktabs}
\usepackage{nicematrix}
\usepackage{todonotes}
\usepackage{physics}
\usepackage{amsthm} 
\usepackage{ bbold }
\usepackage[normalem]{ulem}

\definecolor{forestgreen(web)}{rgb}{0.13, 0.55, 0.13}

\usepackage{cleveref}
\usepackage{lipsum}

\usepackage{algorithm}
\usepackage{algpseudocode}
\usepackage{amsmath}

\algnewcommand{\algorithmicor}{\textbf{ or }}
\algnewcommand{\continue}{\textbf{continue}}
\algnewcommand{\OR}{\algorithmicor}

\makeatletter
\newcommand{\bigcomp}{%
  \DOTSB
  \mathop{\vphantom{\sum}\mathpalette\bigcomp@\relax}%
  \slimits@
}
\newcommand{\bigcomp@}[2]{%
  \begingroup\m@th
  \sbox\z@{$#1\sum$}%
  \setlength{\unitlength}{0.9\dimexpr\ht\z@+\dp\z@}%
  \vcenter{\hbox{%
    \begin{picture}(1,1)
    \bigcomp@linethickness{#1}
    \put(0.5,0.5){\circle{1}}
    \end{picture}%
  }}%
  \endgroup
}
\newcommand{\bigcomp@linethickness}[1]{%
  \linethickness{%
      \ifx#1\displaystyle 2\fontdimen8\textfont\else
      \ifx#1\textstyle 1.65\fontdimen8\textfont\else
      \ifx#1\scriptstyle 1.65\fontdimen8\scriptfont\else
      1.65\fontdimen8\scriptscriptfont\fi\fi\fi 3
  }%
}

\makeatother
\begin{document}

\preprint{AIP/123-QED}

\title{Zero-Noise Extrapolation via Cyclic Permutations of Quantum Circuit Layouts}

\author{Zahar Sayapin}%
 \email{Z.Sayapin@skoltech.ru}
 \affiliation{Skolkovo Institute of Science and Technology, Moscow, Russian Federation}%
\author{Daniil Rabinovich}%
 \email{Daniil.rabinovich@skoltech.ru}
\affiliation{Russian Quantum Center, Moscow, Russian Federation}%
\affiliation{Skolkovo Institute of Science and Technology, Moscow, Russian Federation}%
 \affiliation{Moscow Institute of Physics and Technology, Moscow, Russian Federation}%
\author{Nikita Korolev}%
\affiliation{Russian Quantum Center, Moscow, Russian Federation}%
\author{Kirill Lakhmanskiy}%
\affiliation{Russian Quantum Center, Moscow, Russian Federation}%

\date{\today}

\begin{abstract}
    Increasing the utility of currently available Noisy Intermediate-Scale Quantum (NISQ) devices requires developing efficient methods to mitigate hardware errors. In this work we propose a novel Cyclic Layout Permutations-based Zero-Noise Extrapolation (CLP-ZNE) protocol for such a task. The method leverages the inherent non-uniformity of gate errors in NISQ hardware to extrapolate the expectation value, averaged over cyclic circuit layout permutations, to the level of zero noise. In contrast to the previous layout permutation based approaches, for an $n$-qubit circuit CLP-ZNE requires execution of only $O(n)$ and at most $O(n^2)$ different circuit layouts for circuits of one-dimensional and arbitrary connectivity, respectively. When benchmarked against noise channels modeling the IBM Torino quantum computer, the method reduces a typical error in expectation values of $n=12$ qubit circuits by an order of magnitude, outperforming standard unitary folding ZNE. By demonstrating the ability to mitigate noise of real hardware specifications, including both depolarizing and $T_1/T_2$ relaxation processes, these results give evidence for the applicability of CLP-ZNE to present-day NISQ processors.
\end{abstract}

\maketitle
\section{\label{sec:introduction}Introduction}

Modern day quantum computing is limited to Noisy Intermediate Scale Quantum devices
\cite{preskill2018quantum}. These devices are characterized by limited qubit counts, short coherence times, and imperfect gate operations, which together restrict the depth of quantum circuits that can be executed within a fixed error tolerance \cite{weidenfeller2022scaling,ratcliffe2018scaling,hegde2022toward, mills2022two}. 
Achieving fully fault-tolerant quantum computation ultimately requires the development of quantum error correction~\cite{shor1995scheme}, which currently still faces major experimental and resource challenges.

Facing these limitations, alternative approaches are being actively developed. Among these are variational quantum algorithms (VQAs)\cite{cerezo2021variational, biamonte2021universal, mcclean2016theory, rabinovich2024robustness} which exhibit partial robustness to hardware noise, hardware-efficient methods leveraging native interactions \cite{parra2020digital, rabinovich2022ion, paradezhenko2025heuristic, zhuang2024hardware}, and a variety of quantum error mitigation (QEM) techniques \cite{endo2021hybrid,temme2017error,li2017efficient, hosseinkhani2025noise}. These techniques seek to suppress the impact of noise without the heavy overheads of quantum error correction,  relying instead on modified circuit executions and classical post-processing. QEM encompasses a diverse array of techniques, including, among others, zero-noise extrapolation (ZNE) \cite{temme2017error, endo2018practical, giurgica2020digital, dumitrescu2018cloud, he2020resource, kandala2019error}, probabilistic error cancellation (PEC) \cite{temme2017error, van2023proba}, learning-based protocols~\cite{strikis2021learning, alam2025fermionic}, tensor-network-inspired methods~\cite{fischer2026dynamical}, and symmetry-preserving strategies~\cite{bonet2018low, mcardle2019error, self2024protecting}. A comprehensive survey of the field can be found e.g.~in the review by Cai \textit{et al.}~\cite{cai2023quantum}.

Among these techniques, ZNE stands out for its conceptual simplicity and practically works as a black box. In contrast to approaches such as PEC, this method requires no explicit noise modeling and instead only demands an ability to amplify the noise in a controlled manner. Estimating observables at multiple noise levels subsequently enables extrapolation of the expectation value to the zero-noise limit. Existing noise-amplification strategies include, for example, circuit and gate folding~\cite{giurgica2020digital,dumitrescu2018cloud,he2020resource} as well as pulse stretching~\cite{temme2017error,kandala2019error}. A recent work~\cite{uvarov2024mitigating} proposed to scale noise strength by exploiting spatial variations in gate fidelities. This method takes advantage of non-uniform gate errors by executing the same logical circuit across multiple qubit layouts, thereby effectively scaling the noise. 
Such an approach sits in a sweet spot between noise aware and noise agnostic techniques and can be viewed as a ``gray-box" error-mitigation method. It does not require full noise tomography, but relies on access to gate noise parameters. While this approach has demonstrated notable empirical error suppression, the rigorous guarantees of the method require averaging over all $n!$ qubit permutations, rendering the protocol impractical for large systems.

In this work, we introduce the Cyclic Layout Permutation-based Zero-Noise Extrapolation (CLP-ZNE) protocol. The method preserves the ``gray-box" hardware-awareness of Ref.~\cite{uvarov2024mitigating} while substantially reducing the number of required layouts. For circuits with one-dimensional connectivity, CLP-ZNE requires only $O(n)$ layouts. For arbitrary connectivity, the requirement increases to at most $O(n^2)$.
The method exploits circuit geometry to construct sets of cyclically permuted layouts: linear extrapolation of the corresponding noisy expectation values yields unbiased estimates of noiseless observables up to quadratic noise terms, even under multi-channel noise.

We numerically benchmark CLP-ZNE by estimating the expectation values of random instances of the Sherrington-Kirkpatrick (SK) model, widely used as a mean-field approach to spin glasses~\cite{panchenko2012sherrington,sherrington1975solvable}, and of the transverse-field Ising model minimized by a variational quantum circuit. 
In the numerical simulations, we use the noise models derived from the IBM Torino calibration data \cite{ibm_fake_torino_2025}, exhibiting both depolarizing noise and $T_1/T_2$ relaxation processes. In this setting, for $n=12$ qubit random SK instances, the protocol demonstrates a reduction of typical error by a factor ranging from $8$ to $13$ depending on the protocol adjustment. At the same time, for simplistic noise models, such as depolarization, the error suppression reaches orders of magnitude. 
The method also exhibits significant error suppression under strong noise, and simulations of the transverse-field Ising model further demonstrate its robustness against strong non-unital noise. 
When compared to standard ZNE techniques, the developed approach demonstrated comparable or superior performance. 

The manuscript is structured as follows. Section~\ref{sec:preliminaries} formalizes the multi-channel noise model and derives the perturbative expansion for noisy expectation values. Section~\ref{sec:theory} formulates the CLP-ZNE protocol in the form of two theorems. Section~\ref{sec:numerics} reports numerical simulations of realistic device conditions and quantifies the mitigation performance for the SK model Hamiltonian across different noise settings. A comparison to standard ZNE techniques and a study of the performance of the protocol for high levels of non-unital noise is also presented there. Section~\ref{sec:conclusion} concludes the paper.

\section{\label{sec:preliminaries}Preliminaries}
In this section we establish the formal framework for our analysis. We begin by introducing the necessary definitions and the noise model used to describe errors in quantum circuits. After discussing the generality and interpretational flexibility of the model, we employ first-order perturbation theory to derive analytical expressions for the expectation values of arbitrary observables under the influence of the considered noise.

\subsection{Definitions}

We work within the standard paradigm of gate-based quantum computing, where the universal gate set consists of single and two-qubit gates. For a given two-qubit gate type, the graph of all hardware-supported connections, which may be directed, constitutes the \textbf{device connectivity} (for that gate type). For clarity of the core result of the paper, we restrict our analysis to devices with a single, symmetric two-qubit gate type, e.g., controlled-Z (CZ). In this case, the device connectivity simplifies to a single undirected graph. However, the proposed method can be adapted to non-symmetric two-qubit gates as well. In particular, if such a gate can be implemented in both orientations, the protocol applies directly.

A \textbf{circuit layout} is a mapping from abstract qubits used to define a quantum circuit to the physical qubits of a device on which the circuit is executed. A layout is valid if it employs only the
connections supported by the device hardware.

An \textbf{abstract circuit} is a quantum circuit without specification of the circuit layout, consisting only of operations from the device's universal gate set. We require that any abstract circuit under consideration admits at least one valid layout. If such a layout cannot be found, the circuit must be transformed to obtain one, e.g., by introducing SWAP gates and transpiling these gates into native instructions. An abstract circuit combined with a specific circuit layout constitutes an executable circuit, which is fully specified and ready for execution on the device.

Finally, we define the \textbf{circuit connectivity} as the undirected graph formed by all two-qubit instructions in a given abstract circuit. Distinguishing circuit connectivity from device connectivity is necessary for properly identifying the applications of the developed method to circuits of arbitrary connectivity.

\subsection{Noise model}
We adopt a widely studied noise model in which the errors arise from gate imperfections~\cite{huggins2021virtual, koczor2021dominant}. Let $p$ denote a physical realization of a native two-qubit gate. The action of $p$ on the quantum state is described by a quantum channel $G_{p}$, which could be written as a composition of an intended unitary operation $U_p$ and a residual noise quantum channel $\mathcal{N}_p$:
\begin{equation}
    G_{p} = \mathcal{N}_p \circ U_p, 
    \label{eq:channel}
\end{equation}
which acts on a density matrix $\rho$ as $G_p[\rho]=\mathcal{N}_p[U_p\rho U_p^\dagger]$.
We use a standard Markovian noise model 
\begin{equation}
\label{channel_general_phi}
    \mathcal{N}_p = (1-\sum_{i=1}^d q^i_p)\mathcal{I} + \sum_{i=1}^d q^i_p \Phi_i,
\end{equation}
which represents a combined effect of the identity channel $\mathcal{I}$ and a fixed set of $d$ gate-independent quantum channels that represent gate noise. Gate independence here implies that $\Phi_i$ represent fixed error processes present in all gates, while their strength in specific gate $p$ is controlled by the error rates $q^i_p \geq 0$.  These coefficients depend on the specific qubits involved but are assumed to be independent of the gate's internal parameters. 


Equation~\eqref{channel_general_phi} provides a flexible framework for modeling gate noise. It exactly captures common error channels, including depolarizing, phase-damping, and general Pauli noise. Other sources of noise (such as amplitude damping~\cite{sacchi2017convex, geller2013efficient}, cross-talk, idling errors, and gate miscalibration\cite{rabinovich2024robustness}) can be mapped to an effective Pauli channel using Pauli twirling or randomized compiling.
If noise channels depend on gate parameters (e.g., in the case of parametrized rotations), one can also apply Pauli twirling to reduce them to Pauli noise or append the list $\{\Phi_i\}$ with parameter‑specific processes, at the cost of increasing $d$.

For the convenience of further derivations, we introduce the linear superoperators $\mathcal{E}_i=\Phi_i-\mathcal{I}$, which allow to rewrite channel \eqref{channel_general_phi} in a more compact form
\begin{equation}
\label{channel_general}
    \mathcal{N}_p = \mathcal{I} + \sum_{i=1}^d q^i_p \mathcal{E}_i.
\end{equation}
The superoperators $\{\mathcal{E}_i\}_{i=1}^d$ are a convenient tool, that represents a deviation of quantum channels from the identity, and we will refer to them as noise sources or noise operators later in the text. At small levels of noise, the expression \eqref{channel_general} can also be treated as a linear approximation of an arbitrary channel, thereby accommodating error sources, such as amplitude damping, errors from gate miscalibration, and so on, even without twirling. This brings a physical meaning to artificially introduced superoperators $\mathcal{E}_i$ as derivatives of a parametrized channel with respect to its parameters $q^i$. 
Thus, the model \eqref{channel_general} covers all Markovian processes with the same sources of small errors for all the gates. We further discuss the relationship between operators $\mathcal{E}_i$ and physical quantum channels in Appendix~\ref{sec:appendix_parameters_vs_fidelity}.

\subsection{Influence of noise on the circuit}

The effect of noise on a circuit depends on the choice of the circuit layout. Assume $T$ to be a set of all two-qubit gates in an abstract circuit. Upon a choice of layout $l$ an abstract gate $g \in T$ is mapped to the corresponding noisy physical realization, which we denote $l(g)$, characterized by the error rates $q^i_{l(g)}$. After the application of the noisy circuit, the output state can be written as 
\begin{equation}
    \rho_l =\rho_0+\sum_{s\neq0} a_{l, s} \rho_s,
    \label{scary_full_noise_state}
\end{equation}
where $a_{l, s} = \prod\limits_{i=1}^{d}\prod\limits_{g\in T}(q^i_{l(g)})^{s_g^i}$ and $\rho_0$ is the unperturbed state. This expression represents a mixture of all possible state perturbations $\rho_s$, where the binary matrix $s$ marks the noise operators that were applied after each quantum gate: $s_g^i=1$ ($s_g^i=0$) implies that the noise operator $\mathcal{E}_i$ was (was not) applied  after the ideal gate $g$ in the circuit for $\rho_s$. By construction, the matrices $s$ satisfy the condition $\sum_{i=1}^d s_g^i \le 1$ $\forall g\in T$, implying that only one of the noise operators $\mathcal{E}_i$ might enter upon the application of a single gate as specified in Eq.~\eqref{channel_general}. Note that while the coefficients $a_{l,s}$ explicitly depend on the choice of layout, $\rho_s$ does not depend on it, as $\mathcal{E}_i$ are gate independent. In this formalism, the only role of the layout choice is in changing the relative contributions of each $\rho_s$ by changing the error rates $q^i_{l(g)}$.

Assuming that the error rates $q^i_{l(g)}$ are small and subject to the condition $(\max_{i, g} q^i_{l(g)})|T|d \ll 1$, all quadratic and higher-order terms in $q^i_{l(g)}$ could be neglected. This allows one to obtain a linear approximation of the noisy state $\rho_l$ as

\begin{equation}
    \rho_l =\rho_0 +\sum_{i=1}^d\sum_{g\in T} q^i_{l(g)} \rho_g^i+O(q^2),
    \label{approx_density_matrix}
\end{equation}
where $\rho_g^i=\rho_s$ with $s$ having only one non-zero entry, $s_g^i=1$.
Taking the trace of Eq.~(\ref{approx_density_matrix}) with an observable $H$ we obtain a linear approximation of noisy expectation value $E_l=\operatorname{Tr}\rho_l H$,

\begin{equation}
    E_l= E_0 +\sum_{i=1}^d\sum_{g\in T} q^i_{l(g)} E_g^i+O(q^2),
    \label{energy_with_layout}
\end{equation}
where $E_0=\operatorname{Tr} \rho_0 H$ is the noiseless expectation value and $E_g^i=\operatorname{Tr} \rho_g^i H$ are energy perturbations. 
Evidently, a natural choice of the layout that would minimize the error in the expectation values is the layout with the least noisy gates~\cite{nation2023suppressing}. Yet, as shown in the next section, expression \eqref{energy_with_layout} allows to establish the CLP-ZNE protocol that provides an even better estimate of the noiseless expectation values.

\section{\label{sec:theory}Theory}

In this section we demonstrate that the usage of different circuit layouts allows for executing the zero-noise extrapolation protocol. In contrast to existing proposals, we prove that exact extrapolation (up to terms quadratic in the strength of the noise) can be achieved using only polynomially many permutations. In particular, circuits of one-dimensional connectivity require $O(n)$ different circuit layouts, while arbitrary circuits take at most $O(n^2)$ layout permutations.

\subsection{Use of circuit layouts for error mitigation}
We assume the low-noise regime where Eq.~\eqref{energy_with_layout} is valid. Notice that the perturbed energy terms $E_g^i$ are typically not known and hard to compute, as they depend both on the noise channels and the unperturbed quantum state. Should they be known, the error mitigation would be straightforward, as $E_0$ could easily be deduced after measuring $E_l$. The challenge here is to extract $E_0$ having access only to the noisy expectation values $E_l$ and gate error rates $q^i_{l(g)}$. 

This can be achieved by utilizing a non-uniform distribution of gate infidelities in NISQ devices. As different layouts make use of distinct physical gates, each with their own fidelity, every layout $l$ would produce a distinct expectation value. The authors of Ref. \cite{uvarov2024mitigating} used this observation together with Eq.~\eqref{energy_with_layout}, restricted to a single noise channel, to establish a qubit-permutation-based ZNE protocol.
This approach proposes to extrapolate the dependence of the expectation value $E_l$, computed for random possible circuit layouts $l\in L$, on the circuit error sum $\sum_{g\in T} q_{l(g)}^i$ to the level of zero noise, providing an approximation to the noiseless value. For brevity, we will refer to this protocol as Layout Permutation-based Zero-Noise Extrapolation (LP-ZNE). 

The main drawback of LP-ZNE, however, is the lack of strategy for choosing the layouts: the utility of the protocol is rigorously proven only when all $n!$ layouts are considered, creating immense computational overheads. A potentially less demanding approach could be to perform extrapolation over $q^i_{l(g)}$ themselves, which would require computing energy for $d|T|+1$ circuit layouts, which can still be restrictive due to the factor $|T|$.

\subsection{Cyclic Layout Permutation-based
Zero-Noise Extrapolation}
We propose an alternative approach, which allows to further reduce these computational overheads. To achieve it,  we aim to reduce the number of extrapolation parameters through averaging of the expectation values $E_l$ in Eq.~\eqref{energy_with_layout} over different circuit layouts $l\in L$, yielding
\begin{equation}
\langle E_l \rangle_{l\in L}= E_0 +\sum_{i=1}^d\sum_{g\in T} E_g^i \langle q^i_{l(g)}\rangle_{l\in L}+O(q^2).
    \label{energy_with_layou_avg}
\end{equation}

This is straightforward for circuits with one-dimensional connectivity, where nearest-neighbor gates form a linear or circular topology commonly encountered in practice
\cite{leone2024practical, kandala2017hardware, nakaji2021expressibility}. These circuits can naturally be mapped to cycles $C$ of $m\ge n$ physically connected qubits on the device, with exact equality holding for circuits with cyclic connectivity. More formally, $C$ is a cycle subgraph of the connectivity graph of a considered physical device. Such cycles can typically be identified in devices featuring two-dimensional connectivity, such as superconducting quantum processors~\cite{sdha2025sc, li2025sc}, or in fully connected architectures, as in trapped-ion quantum computers \cite{pogorelov2021compact, bruzewicz2019ion}. Both linear nearest-neighbor and cyclic circuit connectivities are universal, meaning they can be used to execute any quantum algorithm. Furthermore, efficient methods exist for converting an arbitrary circuit connectivity to a linear nearest-neighbor one \cite{saeedi2011synthesis, hirata2009efficient}.

Let $l$ be a layout that maps such a circuit into the qubits and connections of the cycle $C$. Then define $\mathcal{C}(l)$ as a set of all $m$ possible cyclic layout permutations, obtained upon rotations of the assignment $l$ along the qubits of cycle $C$. 
The formulated definitions allow us to state the proposed protocol in the following theorem. To simplify the notation, we will further denote $\langle q_g^i\rangle_{ L}\equiv \langle q_{l'(g)}^i\rangle_{l'\in L}$ and $\langle E\rangle_{ L}\equiv \langle E_{l'}\rangle_{l'\in L}$. 

\vspace{1em}
\textbf{Theorem 1 (Cyclic layout permutation-based zero-noise extrapolation for circuits of one-dimensional connectivity)}\\ 
\textit{Consider an $n$-qubit quantum circuit with one-dimensional connectivity. Assume that the circuit can be embedded into $d+1$ qubit cycles $\{C_j\}_{j=1}^{d+1}$ of a device under a corresponding choice of layouts $\{l_j\}_{j=1}^{d+1}$.
Define $\langle E\rangle_{\mathcal{C}(l_j)}$ as the expectation values averaged over the sets of cyclic layout permutations $\mathcal{C}(l_j)$.
Then a linear extrapolation of  $\langle E\rangle_{\mathcal{C}(l_j)}$ to the level of zero noise with respect to the mean total error rates $\left\langle \sum_{g\in T}q_g^i\right\rangle_{\mathcal{C}(l_j)}$, provides an estimate $E_{\operatorname{mit}}$ that approximates the noiseless expectation value $E_0$  up to terms quadratic in $q$, }
\begin{equation}
    E_{\operatorname{mit}}=E_0+O(q^2).
    \label{eq:theorem_final}
\end{equation}

\begin{proof}

Under a fixed layout $l_j$, each abstract gate $g \in T$ is mapped to a specific physical gate in the cycle $C_j$. Upon cyclic rotations of this layout over the cycle, every abstract gate $g$ is executed exactly once on each of the $m$ physical connections of $C_j$. This makes the mean noise strength $\langle q_g^i \rangle_{\mathcal{C}(l_j)}$ independent of the gate $g$,
\begin{equation}
    \forall g \in T,\left\langle q_g^i\right\rangle_{\mathcal{C}(l_j)}\equiv\frac{\sum_{g' \in T}\left\langle q_{g'}^i\right\rangle_{\mathcal{C}(l_j)}}{|T|} .
    \label{eq:noise_avg}
\end{equation}
Thus, Eq.~\eqref{energy_with_layou_avg} with $L=\mathcal{C}(l_j)$ takes the form
\begin{equation}
    \langle E \rangle_{\mathcal{C}(l_j)}= E_0+\sum_{i=1}^d\Delta_i\left\langle  \sum_{g\in T}q_g^i\right\rangle_{\mathcal{C}(l_j)}+O(q^2),
    \label{eq:lemma1_output}
\end{equation}
where $\Delta_i=\left(\sum_{g\in T} E_g^i\right)/|T|$ is the average perturbation originating from the error source $\mathcal{E}_i$.

Introducing a design matrix $X$ \cite{Timm2002} with elements
    \begin{equation}
        X_{ji} = 
        \begin{cases}
            ~~~~~1, ~~~~~~~~~~~~~~~~~~~i=0\\
            \left\langle  \sum_{g\in T}q_g^i\right\rangle_{\mathcal{C}(l_j)}, ~~1\le i\le d,
        \end{cases}
        \label{eq:design_matrix}
    \end{equation}
    Eq.~\eqref{eq:lemma1_output} for $1\le j\le d+1$ can be rewritten as $\langle E\rangle_{\mathcal{C}(l_j)} =  X_{j0} E_0 + \sum\limits_{i=1}^d X_{ji}\Delta_i+O(q^2)$. 
    This dependence can be fit with a linear model which, after minimizing the square distance to the noisy expectation values, yields the mitigated expectation 
\begin{equation}
    E_{\operatorname{mit}} = \sum_{j=1}^{d+1} (X^{-1})_{0j} \langle E \rangle_{\mathcal{C}(l_j)}.
\end{equation}
    Finally, we note that the first row of $X^{-1}$ remains finite for $q\to0$, implying that $E_{\operatorname{mit}}= E_0+\sum\limits_{j=1}^{d+1}(X^{-1})_{0j}O(q^2) = E_0+O(q^2)$, which finishes the proof.   
\end{proof}

\vspace{1em}
Several remarks should be made concerning the proof of Theorem 1.
First, the embeddings of circuits with linear and circular connectivities into cycles of connected qubits were considered only out of convenience. This choice allows to generate cyclic layout permutations $L=\mathcal{C}(l_j)$, which automatically provide gate independence of $\langle q_g^i \rangle_L$. The theorem, however, remains valid for any sets of permutations, provided that this condition is fulfilled.

Second, the proof uses the invertibility of the matrix $X$, i.e.~that the associated vectors of noise strengths are linearly independent. This condition is typically straightforward to satisfy in practice, given the inherently non-uniform fidelity distributions present in NISQ devices. In cases where the chosen permutation cycles have led to linearly dependent columns of $X$, extra qubit cycles can be considered to append rows to $X$ until the linear independence of the columns is achieved. Once this is fulfilled, solution for $E_0$ can be extracted using the Moore–Penrose pseudoinverse $(X^{\top}X)^{-1}X^{\top}$ instead of $X^{-1}$.

Finally, Theorem 1 also holds in the presence of single-qubit gate errors, even when the corresponding noisy channels differ from those of the two-qubit gates. The only additional requirement in this case is the inclusion of more layout permutations to ensure full rank of $X$.

The conditions and proof of Theorem 1 thus establish the essential theoretical foundations of the proposed CLP-ZNE protocol. Below we summarize its main steps as an algorithm for circuits of a one-dimensional topology and graphically illustrate the required steps in Fig.~\ref{scheme_CLP}.
\addtocounter{algorithm}{-1} 
\renewcommand{\thealgorithm}{\relax} 
\begin{algorithm}[H]
\caption{CLP-ZNE (Cyclic Layout Permutation-based Zero-Noise Extrapolation)}
\begin{algorithmic}[1]
\Require Quantum circuit with one-dimensional connectivity $Q$ on $n$ qubits, observable $H$ 
\Ensure Mitigated expectation value of the observable $E_{\operatorname{mit}}$
\State Select $d+1$ qubit cycles $\{C_j\}_{j=1}^{d+1}$ with length $n$ for cyclic circuit connectivity and any $m\ge n$ for linear connectivity.  
\State For each qubit cycle $C_j$ choose any layout $l_j$ which maps the abstract circuit to the qubits and connections of the corresponding cycle.
\For{$j = 1$ to $d+1$}
    \State Generate a set of cyclically permuted layouts $\mathcal{C}(l_j)$ 
    \State For each layout $l' \in \mathcal{C}(l_j)$, compute the expectation value of observable $E_{l'}=\operatorname{Tr}\rho_{l'} H$, where $\rho_{l'}$ is the state prepared on a quantum computer by executing $Q$ using layout $l'$
    \State Average the computed observables: $\langle E\rangle_{\mathcal{C}(l_j)} = \text{average}( \{E_{l'} \mid l' \in \mathcal{C}(l_j)\} )$
    \State Calculate average total error rates $ e^i_j =\left\langle  \sum_{g\in T} q_g^i\right\rangle_{\mathcal{C}(l_j)}\equiv\left\langle  \sum_{g\in T} q_{l'(g)}^i\right\rangle_{l'\in\mathcal{C}(l_j)}$
\EndFor
\State Linearly extrapolate $\langle E\rangle_{\mathcal{C}(l_j)}$ over the average total error rates $e^i_j$ to the zero-noise limit to obtain $E_{\operatorname{mit}}$
\State \Return $E_{\operatorname{mit}}$
\end{algorithmic}
\end{algorithm}

\vspace{1.5em}

The developed protocol can be generalized to quantum circuits beyond just linear connectivity. The key element, again, would be to identify layout permutations that ensure gate independence of certain average noise strengths $\langle q_g^i \rangle_L$. For circuits with arbitrary connectivity, this can still be achieved using cyclic layout permutation, yet the approach would require all-to-all connectivity on the quantum hardware. Assuming this setting, we formulate the CLP-ZNE protocol for quantum circuits with arbitrary entangling-gate connectivities in the following theorem.

\textbf{Theorem 2 (Cyclic layout permutation-based zero-noise extrapolation for general circuit connectivity)} \\
\textit{Consider an $n$-qubit quantum circuit implemented on a device with an all-to-all connectivity. There exist a set of $M=Kd+1$ layouts $\{l_j\}_{j=1}^{M}$ mapping the circuit onto corresponding qubit cycles $\{C_j\}_{j=1}^{M}$ and a partitioning of the circuit gates into $K$ gate types $T = \bigsqcup_{k=1}^K T^k$, with
\begin{itemize}
    \item $K=2$ for square-grid circuit connectivity;
    \item $K \le  \lfloor \frac{n}{2} \rfloor$ for arbitrary circuit connectivity,
\end{itemize}
which allow executing zero-noise extrapolation. Define $\langle E\rangle_{\mathcal{C}(l_j)}$ as the expectation values averaged over the sets of cyclic layout permutations $\mathcal{C}(l_j)$. 
Then a linear extrapolation of  $\langle E\rangle_{\mathcal{C}(l_j)}$ to the level of zero noise with respect to the mean total error rates $\left\langle \sum_{g\in T^k}q_g^i\right\rangle_{\mathcal{C}(l_j)}$, provides an estimate $E_{\operatorname{mit}}$ that approximates the noiseless expectation value $E_0$  up to terms quadratic in $q$, }
\begin{equation}
    E_{\operatorname{mit}}=E_0+O(q^2).
    \label{eq:theorem_unified}
\end{equation}
The proof of this theorem largely follows the steps of Theorem 1, but requires a specific construction of circuit layouts. We present the proof of Theorem 2 in Appendix \ref{sec: proofs}. In Appendix \ref{bias_estimate} we also estimate the CLP-ZNE quadratic bias of Eqs. \eqref{eq:theorem_final} and \eqref{eq:theorem_unified} arising from truncating perturbative noise expansion.

We summarize the requirements for the proposed CLP-ZNE protocol in Table~\ref{summary}. The computational overhead is expressed as the total number of distinct layouts required for noisy observable estimation, 
which contributes to the total execution time.
Most importantly, this overhead scales either as $O(n)$ or $O(n^2)$ depending on the circuit connectivity. This illustrates that while arbitrary connectivities necessitate a larger set of layout permutations, the protocol remains computationally tractable for NISQ-era devices.

While CLP-ZNE can provably mitigate errors arising from gate imperfections, it does not account for other noise sources such as state preparation and measurement (SPAM) errors. Thus, it has to be used in conjunction with dedicated SPAM error mitigation techniques. Similarly, shot noise in measurement statistics would also adversely affect the protocol's performance. In Appendix~\ref{sec:shot noise}, we estimate its impact on the error mitigation accuracy.

\begin{widetext}

\begin{figure*}[h]
\centering
\includegraphics[width=\textwidth]
{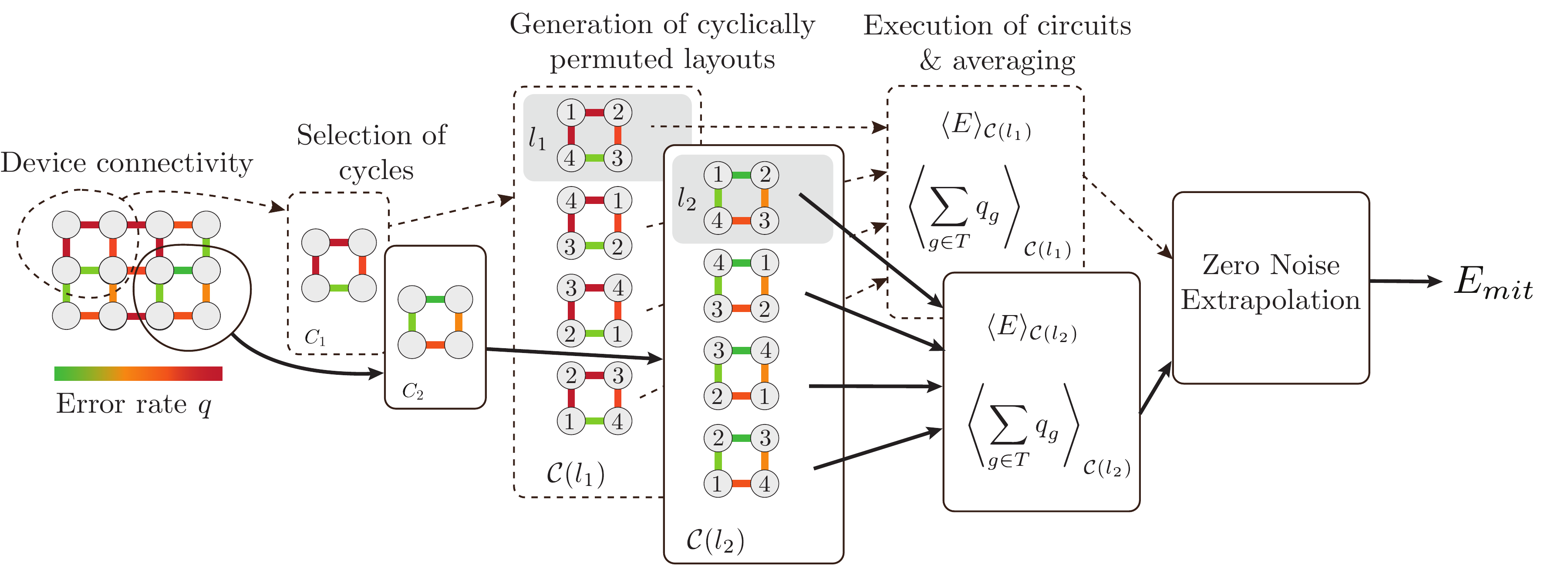}
\caption{Schematic illustration of the proposed CLP‑ZNE protocol applied to a four‑qubit circuit under single‑parameter noise ($d=1$). Circles represent physical qubits, while edges denote available two-qubit couplings, color-coded by their individual error rates $q$. The procedure begins by selecting two qubit cycles, $C_1$ and $C_2$, chosen such that their total error rates differ. For each cycle, an initial layout ($l_1$ for $C_1$, $l_2$ for $C_2$) is selected to map the abstract circuit onto the physical qubits and couplings; the numerical labels indicate the assignment of abstract qubits to physical positions. Subsequently, cyclically permuted layout sets $\mathcal{C}(l_1)$ and $\mathcal{C}(l_2)$ are generated by rotating each initial layout along its corresponding cycle. For every layout within these sets, both the observable expectation value $\langle E \rangle$ and the cumulative error rate $\langle \sum_{g \in T} q_g \rangle$ are computed and then averaged over their respective sets. Finally, the averaged noisy values are linearly extrapolated to the zero-noise limit, yielding the error-mitigated expectation value $E_{\text{mit}}$.}
\label{scheme_CLP}
\end{figure*}

\begin{table*}[h]
\centering

\begin{tblr}{
  width = \textwidth,
  colspec = {|X[l]|X[c]|X[c]|X[c, 2.5cm]|X[c, 2.5cm]|}, 
  hlines, vlines, 
}
  \SetCell[r=2]{l} CLP-ZNE requirements & \SetCell[c=4]{c} Circuit connectivity& & \\
  & Linear & Cyclic & Square-grid & Arbitrary \\
  Device connectivity & {Any containing $d+1$ \\cycles of $m\ge n$ qubits}&{Any containing $d+1$ \\cycles of $n$ qubits}& All-to-all&All-to-all\\
  Number of layouts & $m(d+1)$& $n(d+1)$&$n(2d+1)$& $\le n\left(\left\lfloor\frac{n}{2}\right\rfloor d+1\right)$\\
\end{tblr}
\caption{Summary of CLP‑ZNE requirements for an $n$‑qubit quantum circuit under a $d$‑parameter noise model, shown for different circuit connectivities}
\label{summary}

\end{table*}
\end{widetext}

\section{\label{sec:numerics}Numerical verification}
In this section we numerically validate the protocol for realistic noise models, derived from existing quantum hardware, that account for both single- and two-qubit gate errors. The simulations are conducted for both one-dimensional and square grid circuit connectivities. 

For circuits of cyclic connectivity the simulations are performed with the Qiskit framework~\cite{qiskit} using the \textsc{FakeTorino} backend, following the characteristics of the IBM Torino quantum computer \cite{ibm_fake_torino_2025} utilizing Heron architecture \cite{abughanem2025heron}. This device is a natural choice, as it features symmetric entangling CZ gates, assisting CLP-ZNE implementation. 
Circuits with square-grid topology are executed using \textsc{NClusterBackend}, a synthetic custom Qiskit backend designed to model a processor composed of independent qubit clusters with all-to-all intra-cluster connectivity. This backend was specifically constructed to accommodate the requirements of CLP-ZNE for quantum circuits with nonlinear connectivity. To ensure realistic noise conditions, \textsc{NClusterBackend} parameters are initialized by sampling qubit properties ($T_1$, $T_2$) and gate errors from normal distributions, with the statistical means and standard deviations extracted from \textsc{FakeTorino} backend.

To perform CLP-ZNE, we assume access to the noise parameters $q^i_p$ of the considered device. If the error is dominated by a single incoherent channel ($d = 1$), the average infidelity $1 - F_{\operatorname{avg}}$ of a physical gate $p$ is directly proportional to the parameter $q_p \equiv q^{1}_p$ within the linear approximation of the noisy channel, as shown in Appendix~\ref{sec:appendix_parameters_vs_fidelity}. In this case, the mitigation can be performed directly over the average infidelity since the proportionality factor does not affect the extrapolation. The average gate fidelity $F_{\operatorname{avg}}$ can be obtained by means of  randomized benchmarking \cite{knill2008bench, silva2025bench} and is typically known from the device specification. Interestingly, even in the multichannel case ($d > 1$), extrapolation over the gate infidelities still leads to a noticeable reduction in error, as demonstrated below. Nevertheless, a complete implementation of the protocol would, in principle, require multidimensional extrapolation over all relevant error sources.

\subsection{Simulations for realistic noise} 
To demonstrate the ability of CLP-ZNE to suppress the errors in expectation values of measured observables, we compare the distributions of errors obtained from noisy circuit simulations before and after error mitigation. We use 12-qubit and 9-qubit \texttt{TwoLocal} ansatzes \cite{qiskit} with cyclic and square-grid connectivity, respectively. Both ansatzes consist of single-qubit rotations ($R_x$ and $R_z$) and 12 entangling CZ gates per layer. For each case, 20 random 3-layered circuits are generated by initializing circuit parameters uniformly on the interval $[0, 2\pi)$. After transpilation, 12-qubit circuits had 36 CZ, 96 $\sqrt{X}$ gates (the square root of the Pauli $X$), and 108 $R_z$ gates; 9-qubit circuits had 36 CZ, 72 $\sqrt{X}$ gates, and 81 $R_z$ gates. For both cyclic and square-grid connectivity types we generate 100 instances of the Sherrington–Kirkpatrick (SK) \cite{panchenko2012sherrington,sherrington1975solvable} model
\begin{equation}
    H_{\operatorname{SK}} =\frac{1}{\sqrt{n}} \sum_{i \neq j} J_{ij} Z_i Z_j+h\sum_{i=1}^{n} X_i,
    \label{sk_model}
\end{equation}
of the corresponding system sizes. Here $J_{ij} \sim\mathcal{N}(0, 1)$, $Z$ and $X$ are the Pauli operators and the transverse field strength $h$ is set to $1$. To ensure consistency, each observable was evaluated at every circuit of the corresponding type.
The full density matrix simulations are implemented to obtain noisy expectation values of the generated SK instances in different circuit layouts, which are then used for extrapolation.

\begin{figure}[tbh]
    \includegraphics[width=0.5\textwidth]{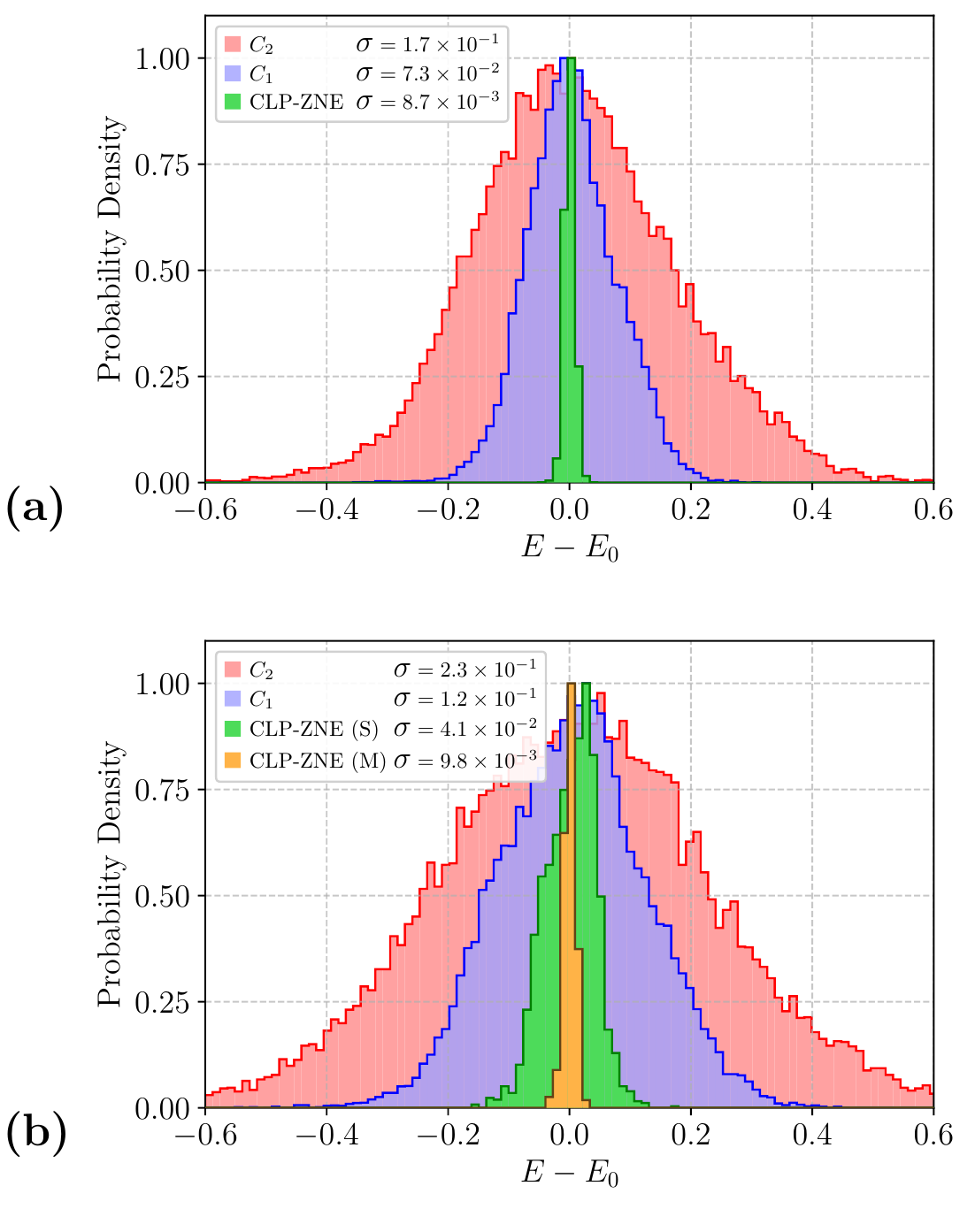}
            \caption{The distributions of errors before and after the mitigation using CLP-ZNE for circuits of cyclic connectivity. (a) Noise model derived from IBM Torino QPU calibration data; (b) analogous noise model, where each $T_1$, $T_2$ is reduced by a factor of 10. Panel (b) compares two versions of the protocol: a single-parameter extrapolation over 2 cycles (S) and a multi-parameter extrapolation over 4 cycles (M). Cycle $C_1$ corresponds to the least noisy cycle on the device, while $C_2$ has approximately double the infidelity. $\sigma$ represents standard deviations of the corresponding distributions. The distributions are normalized to the same height to facilitate visual comparison.}
            \label{fig:histograms}
\end{figure}

\begin{figure}[h]
    \centering
\includegraphics[width=\linewidth]{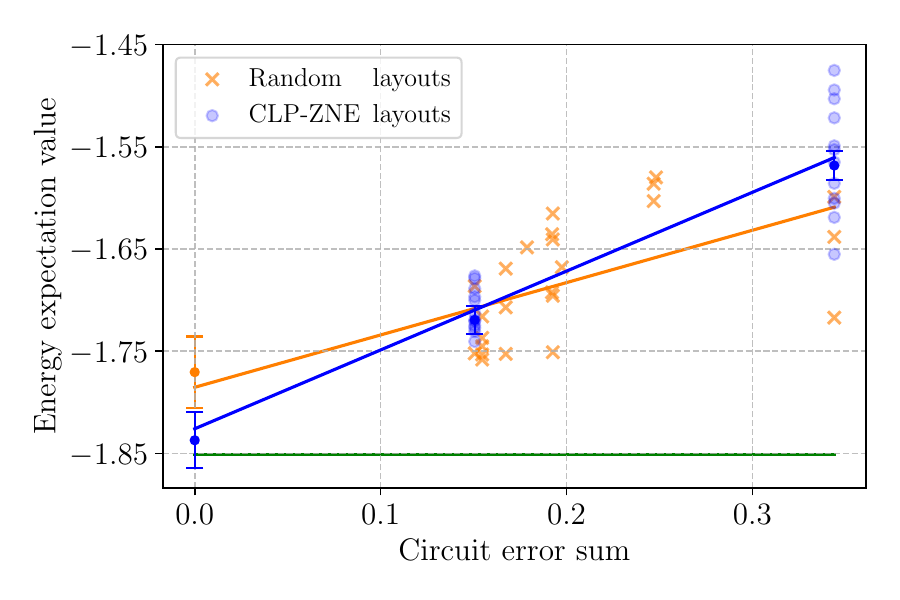}
   \caption{Representative example of CLP-ZNE (blue) and LP-ZNE with random layout permutations (orange) for a specific $n=12$ qubit instance from Fig.~\hyperref[fig:histograms]{2(a)} in the presence of shot noise. Both methods utilize $2n=24$ circuit layouts. Orange crosses and transparent blue circles denote energy expectation values obtained with $10^4$ shots per layout and measurement basis. Solid circles indicate the exact mean energy values $\langle E\rangle_{\mathcal{C}(l_j)}$
  and the mitigated values 
$E_{\operatorname{mit}}$ computed without sampling noise, with error bars showing the theoretical standard deviation expected from finite-shot measurements. The horizontal green line denotes the exact noiseless energy.}
    \label{fig:ZNE_example}
\end{figure}

The results for circuits of cyclic connectivity are summarized in Fig.~\ref{fig:histograms}, which demonstrates the distributions of errors of estimated expectation values $E_l-E_0, l\in{\mathcal{C}(l_j)}$, obtained from cyclic permutations of several layouts $l_j$. It is compared to the distribution of errors of mitigated values $E_{\text{mit}}-E_0$, obtained from the CLP-ZNE protocol, conducted with respect to the total circuit infidelity.
Figure \hyperref[fig:histograms]{2(a)} demonstrates the results for the noise specification of the IBM Torino device, exhibiting depolarization as well as amplitude damping and dephasing originating from $T_1/T_2$ noise. The difference in $E-E_0$ distributions for different cycles clearly indicates the influence of noise inhomogeneity on the measurement data. It is seen that the CLP-ZNE protocol suppresses the typical scale of energy error by a factor of 8 when compared to the layouts on the least noisy cycle $C_1$, and provides even more dramatic improvement when compared to layouts on noisier cycle $C_2$.  This confirms the applicability of the linear expansion \eqref{approx_density_matrix} and the single-channel $d=1$ approximation for CLP-ZNE realization on the considered NISQ processor. The latter is further verified by a comparison of the channels' amplitudes: depolarization contributes $\sim 0.13$ to the total circuit infidelity of $0.15$ on cycle $C_1$ (see Fig.~\ref{fig:ZNE_example}), while $T_{1}/T_2$ processes only contribute the remaining $0.02$. Apart from single-channel approximation, the remaining width of the distribution is also attributed to the effect of single-qubit gate errors. In their absence, the width of the distribution reduces by a factor of $13$, providing even stronger error suppression. 
Moreover, in a simplified noise model in which all gate errors stem solely from depolarizing noise with an average gate infidelity of approximately $10^{-3}$, we observe a reduction in the characteristic error scale by several orders of magnitude.

To test the versatility of the proposed protocol, we artificially reduce $T_{1},T_2$ times for every qubit by a factor of 10, which increases the relative contribution of these processes to the overall noise profile. The resultant noise model has the least noisy cycle infidelity $\sim 0.23$ with negligible input from depolarization. 
In this setting, due to comparable contributions from damping and pure dephasing channels, the quality of single-parameter extrapolation degrades, as demonstrated in green in Fig.~\hyperref[fig:histograms]{2(b)}. Yet the protocol still reduces typical error by a factor of $3$ compared to the least noisy circuit layouts. The performance of the protocol can be further improved in this case by performing a multidimensional extrapolation, as per protocol requirements, over $d+1 = 4$ qubit cycles (48 circuit layouts), as shown in orange in Fig.~\hyperref[fig:histograms]{2(b)}. In this case CLP-ZNE suppresses the error by a factor of $9$ compared to the cycle $C_1$. Overall, these simulations demonstrate the CLP-ZNE protocol utility in various practically relevant conditions, expanding its applicability range compared to the analytical expectations.

\begin{figure}[tbh]
    \includegraphics[width=0.48\textwidth]{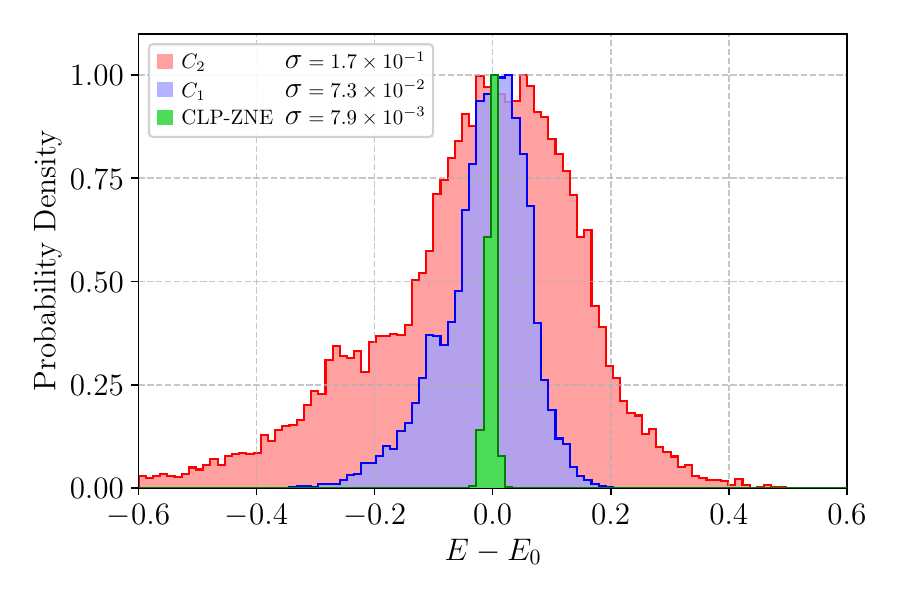}
            \caption{The distributions of errors before and after the mitigation using CLP-ZNE for circuits of square-grid connectivity, for the noise model from \textsc{NClusterBackend}. The extrapolation is performed over gate infidelities, i.e.~using single-channel approximation ($d=1$) and $dK+1 = 3$ qubit cycles (27 layouts). Cycle $C_1$ corresponds to the least noisy cycle on the device, while $C_2$ has approximately double the infidelity.}
            \label{fig:histograms^2}
            
\end{figure}

Circuits of square grid connectivity demonstrate comparable results. Figure~\ref{fig:histograms^2} depicts the error distributions obtained before and after the mitigation for the noise parameters of \textsc{NClusterBackend}. In this case CLP-ZNE suppresses the typical error by a factor of 9 compared to least noisy cycle $C_1$.
\vspace{-1em}

\vspace{-1em}
In Fig.~\ref{fig:ZNE_example}, we exemplify a typical realization of the CLP-ZNE protocol for the same settings as in Fig.~\hyperref[fig:histograms]{2(a)}. For illustration, the simulation was conducted with finite statistics, using $10^4$ shots per $X$ and $Z$ bases for measuring the observable Eq.~\eqref{sk_model} in every circuit layout, demonstrating the stability of the mitigated value with respect to finite statistics. Detailed analytical estimates on the effect of statistical error on the error-mitigation efficiency are presented in Appendix~\ref{sec:shot noise}. The result is also contrasted with the LP-ZNE protocol, with extrapolation performed over the same number of circuit layouts chosen randomly.

\subsection{Benchmarking}
To evaluate the robustness of CLP-ZNE across varying hardware architectures and noise regimes, we benchmark its performance against three established error mitigation approaches: LP-ZNE with random layout permutations, circuit-folding ZNE and gate-folding ZNE \cite{majumdar2023best}. The numerical evaluation consists of two series of simulations. The first series utilizes 100 random Sherrington-Kirkpatrick (SK) model instances \eqref{sk_model}, testing each observable on 20 distinct 12-qubit circuits with cyclic connectivity. The second series employs a separate set of 100 SK instances, testing each observable on 20 distinct 9-qubit circuits with square-grid connectivity. All circuits were constructed using 3-layer \texttt{TwoLocal} ansatzes~\cite{qiskit}, comprising single-qubit rotations ($R_x$ and $R_z$) and CZ entangling gates, with the parameters sampled uniformly from $[0, 2\pi)$. All results in this subsection were obtained via full density matrix simulations.

Table~\ref{tab:benchmarking results} presents the median error suppression factors $|E_{\text{noisy}} - E_0|/|E_{\text{mit}} - E_0|$ for each method under (i) the original noise conditions, (ii) with $T_1/T_2$ noise increased by a factor of 10, and (iii) non-symmetric local Pauli noise with entangling gate infidelities comparable to the \textsc{FakeTorino} setting and ideal local operations. For the original noise setting, CLP-ZNE was implemented using single-parameter extrapolation, i.e.~assuming $d=1$ and using $2$ cycles for circuits with cyclic connectivity (24 layouts per circuit instance) and 3 cycles for circuits with square-grid topology (27 layouts per circuit instance). For the increased $T_1/T_2$ noise setting and Pauli noise, multi-parameter extrapolation ($d=3$) was employed, utilizing 4 cycles for cyclic topologies (48 layouts) and 7 cycles for square-grid topologies (63 layouts). The LP-ZNE protocol was implemented using single-parameter extrapolation with the same number of layouts as for CLP-ZNE in each corresponding setting, which were selected at random. Both gate-folding and circuit-folding ZNE were evaluated using 4-point linear extrapolation (L) and exponential fit (E) with noise amplification factors $\{1, 3, 5, 7\}$ \cite{majumdar2023best, cai2023quantum}.

Under original noise conditions, CLP-ZNE achieves median error suppression factors of 8.4 (\textsc{FakeTorino}) and 9.6 (\textsc{NClusterBackend}). This substantially outperforms LP-ZNE ($2.0-4.2$) and most unitary folding ZNE approaches ($1.2-2.9$), losing only to gate folding with an exponential fit. When $T_1/T_2$ noise is increased tenfold, CLP-ZNE maintains superior performance with suppression factors of 12.8 and 7.4, respectively, even outperforming gate folding with the exponential fit. Similar behavior is also observed for non-symmetric Pauli noise. Interestingly, as the amplitude of the Pauli noise decreases, CLP-ZNE starts to demonstrate a higher error suppression factor compared to gate folding with an exponential fit. Overall, the reported interquartile ranges indicate that CLP-ZNE's advantage persists across different circuit instances and noise conditions. It is also worth noting that CLP-ZNE outperforms LP-ZNE even when the latter uses more permutations. For example, for 
\textsc{NClusterBackend} LP-ZNE performance improves from $4.2$ to just $5.3$ and $7.5$ when using $108$ and $432$ permutations, respectively.

\begin{widetext}
 
\begin{table*}[h]
\centering
\begin{tblr}{
  width = \textwidth,
  colspec = {|p{2.4cm}
  |X[c]|X[c]|X[c]|X[c]|}, 
  hlines, vlines, 
}
  \SetCell[r=2]{l} Noise type & \SetCell[c=4]{c} Error suppression factor $|E_{\text{noisy}}-E_0|/|E_{\text{mit}}-E_0|$& & \\
  & {CLP-ZNE\\ (this paper)} &{LP-ZNE\\(random layouts)}& {ZNE \\ (circuit folding)} & {ZNE \\ (gate folding)} \\
  \SetCell[c=5]{c} FakeTorino, 12-qubit circuits with cyclic connectivity & & & \\
  Original noise & 8.4 [5.1–12.4]& 2.0 [0.7–4.9]& {2.9 [1.5–5.3] (L)\\2.3 [1.0–5.0] (E)}&{2.6 [2.0–3.4]  (L)\\\textbf{13.4 [8.0–18.3] (E)}}\\
  {$T_1/T_2$ noise \\increased} $\times 10$ & \textbf{12.8 [6.2–26.7]} & 1.6 [0.9–3.1]& {1.7 [1.0–3.0] (L)\\1.4 [0.6–2.8] (E)} &{ 1.9 [1.4–2.4] (L)\\
  5.0 [3.1–9.2] (E)}\\
  Pauli noise & \textbf{40.5 [25.5–60.2]} & 3.6 [1.5–8.7]&{2.3 [0.9–4.9] (L)\\1.3 [0.6–2.9](E)}&{5.3 [3.4–8.1] (L)\\12.5 [5.4–25.8](E)}\\
  \SetCell[c=5]{c} NClusterBackend, 9-qubit circuits with square-grid connectivity ($3 \times 3$) && & & \\
  Original noise & 9.6 [4.6–16.6] &4.2 [2.1–11.3]&{ 2.2 [1.1–4.5] (L)\\1.2 [0.5–2.5] (E)}& {2.1 [1.7–2.7] (L) \\\textbf{18.0 [11.5–28.8]} (E)}\\
  {$T_1/T_2$ noise \\increased $\times 10$} & \textbf{7.4 [5.3–10.6]}&1.6 [1.0–2.7]&{1.4 [1.0–2.1]  (L)\\0.9 [0.4–1.9] (E)} &{1.6 [1.3–2.0] (L) \\ 5.8 [3.3–9.6] (E)}\\
  Pauli noise & \textbf{17.6 [12.0–25.0]} & 3.9 [1.8–7.5] 
  & {1.3 [0.6–3.1](L)\\0.8 [0.4–1.9] (E)} & {3.9 [2.9–5.9] (L)\\\textbf{17.6 [8.2–37.6] (E)}}
\end{tblr}
\caption{Comparison of error mitigation methods across backends, noise settings, and circuit connectivities. Values are reported as median and [Q1--Q3], where higher error suppression factors $|E_{\text{noisy}} - E_0|/|E_{\text{mit}} - E_0|$ indicate better performance. For each method, $E_{\text{noisy}}$ is evaluated using the same layout with the smallest total infidelity. 
LP-ZNE uses the same number of layouts as CLP-ZNE in each noise setting. Both gate-folding and circuit-folding ZNE are evaluated using 4-point linear (L) and exponential (E) extrapolation with noise amplification factors $\{1, 3, 5, 7\}$.} 

\label{tab:benchmarking results}
\end{table*}
\end{widetext}

\subsection{Simulations for strong non-unital noise} 
In this section we investigate the role of high-order terms in Eq.~\eqref{approx_density_matrix} by studying how the protocol performance degrades upon noise strength increase. We assume that the entangling two-qubit gates are subject only to the non-unital amplitude damping channels \cite{schuster2024, PhysRevA.105.042406}, acting on both qubits separately, with the associated non-uniform coherence times $T_1$. Note that in the case of amplitude damping, the form \eqref{channel_general} only serves as a linear expansion of the channel, with $\mathcal{E}$ originating from the derivative of the channel with respect to its strength.
While, in theory, the protocol performance should not depend on the noise type, non-unital noise is known to nontrivially change the performance of quantum algorithms \cite{fefferman2024effect, wang2024can, fontana2022non}, making it a challenging testbed for error mitigation implementation.  
Ultimately, in this section we do not aim at replicating realistic device parameters, but test the robustness of the protocol against the non-unital noise strength increase. 

While the amplitude damping model can readily be built in terms of $T_1$ time, here we implement it directly in terms of the channel strength $\gamma = 1 - e^{-t/T_1}$ ($t$ being the gate duration) \cite{PhysRevA.105.042406},
which enters the channel via the Kraus operators
\begin{equation}
    K_0 = \begin{pmatrix} 1 & 0 \\ 0 & \sqrt{1 - \gamma} \end{pmatrix}, \quad 
K_1 = \begin{pmatrix} 0 & \sqrt{\gamma} \\ 0 & 0 \end{pmatrix}.
\end{equation}

We assume this single-qubit error channel is applied after each entangling gate in the circuit to both qubits with the associated strengths.
To specify the problem, we focus on preparing the ground state of the cyclic transverse-field Ising  Hamiltonian
\begin{equation}
    H_I = \sum_{j=1}^{n} Z_j Z_{j+1} + \sum_{j=1}^{n} X_i,
\end{equation}
with $Z_{n+1} \equiv Z_1$ by means of the Variational Quantum Eigensolver (VQE)  \cite{mcclean2016theory, tilly2022vqe} circuit. We use an $n=12$ qubit $p=4$ layer \texttt{TwoLocal} ansatz \cite{qiskit}, with each layer comprising a block of single-qubit \(R_y(\theta)\) rotations followed by a cyclic nearest-neighbor entangling layer of controlled-NOT (CNOT) gates. In the absence of noise the circuit optimization provides VQE energy within the energy error of $0.3$ from the true ground state. 

\begin{figure}
\includegraphics[width=8cm]
{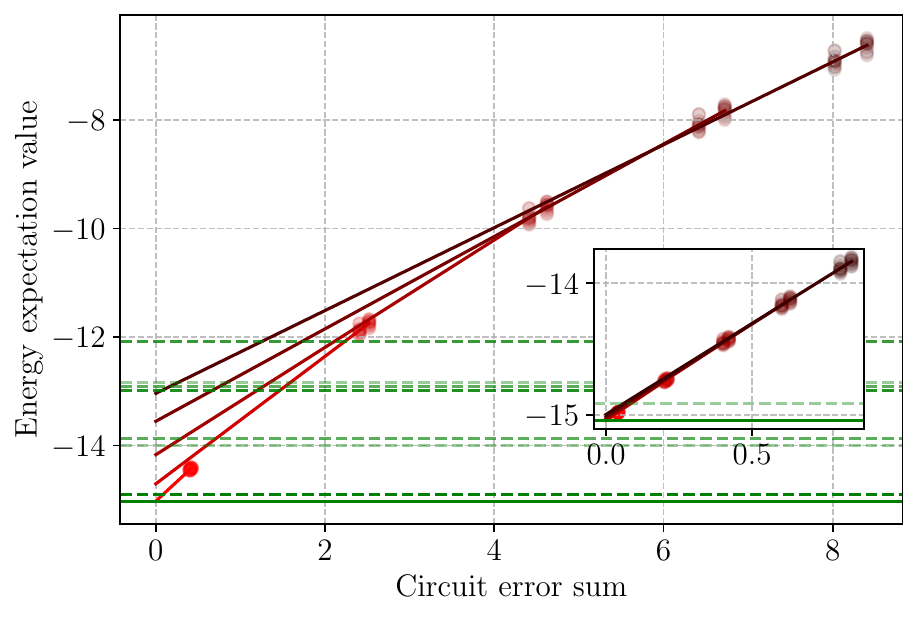}
\caption{CLP-ZNE realization for rescaled amplitude damping noise for 12 qubit VQE circuit. Different clusters of points depict noisy expectation values, with darker colors corresponding to noisier circuits. Solid horizontal line depicts the noiseless VQE energy, while the dashed lines are elevated by the energies of lowest excitations of $H_I$. Inset: Same plot in the range of small circuit errors.}
\label{fig:exp_energy_vs_t1_noise}
\end{figure}

\begin{figure}
\includegraphics[width=8cm]
{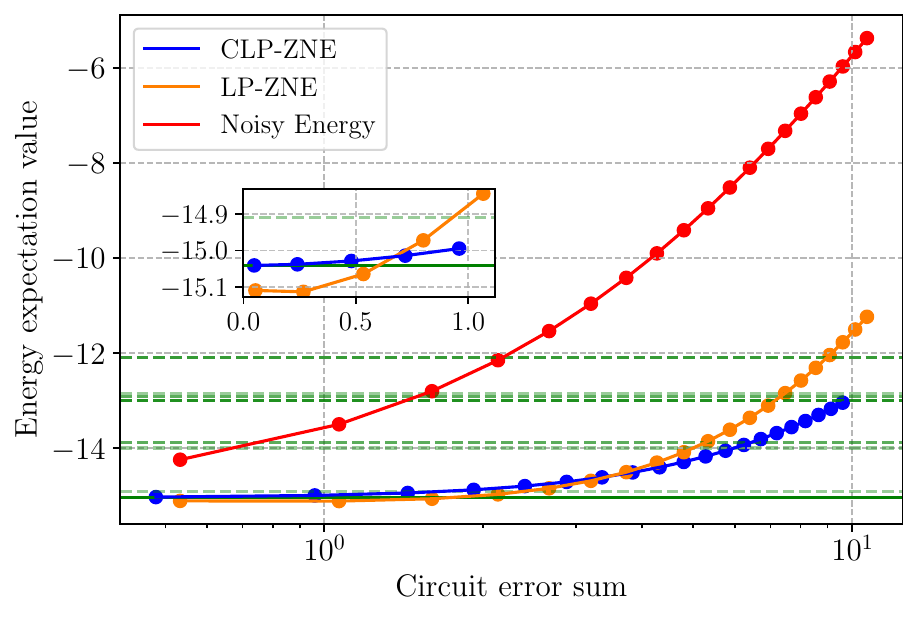}
\caption{Noisy and mitigated energies for both CLP-ZNE and LP-ZNE obtained for different circuit error sums $2p\sum_i\gamma_i$ by rescaling the strengths $\gamma_i$. Solid horizontal line depicts the noiseless VQE energy, while the dashed lines are elevated by the energies of lowest excitations of $H_I$. Inset: Same plot in the range of small circuit errors.}
\label{fig:mit_energy_vs_t1_noise}
\end{figure}

After the training, the circuit is subject to noise as explained above and CLP-ZNE is performed. For this we produce two sets of 12 values of $\gamma_i^0$, which are obtained via sampling $T_1$ times from the IBM Torino calibration data, with gate duration $t = 75$ ns. The CLP-ZNE protocol is then executed multiple times for noise strengths $\gamma_i = \lambda \gamma_i^0$. Such uniform noise scaling allows to directly assess protocol performance with respect to noise strength, without altering the relative contributions of different channels. For each rescaling, CLP-ZNE is then conducted with respect to the circuit error sum $2p\sum_i \gamma_i$, proportional to the sum of circuit gate infidelities. Note that although every entangling gate is now followed by two noisy channels, single-parameter $d=1$ extrapolation is still justified due to the structure of the circuit, as can be confirmed with Eq.~\eqref{eq:lemma1_output}. The results of the simulations are demonstrated in Fig.~\ref{fig:exp_energy_vs_t1_noise}. It is seen that while for the low noise strength the extrapolation accurately predicts the noiseless VQE, even within the gap of Hamiltonian $H_I$, the performance clearly degrades upon noise increase by deviating from the linear trend. For large circuit error sums the extrapolation overestimates the noiseless energy, however, the energy error still reduces by about a factor of $2.5$.
A direct comparison of the noisy and mitigated energy is demonstrated in Fig.~\ref{fig:mit_energy_vs_t1_noise}. The mitigated expectation value is seen to be within a spectral gap of $H_I$ from the noiseless VQE for circuit error sum $2p \sum\gamma_i\lesssim 1$, comparable to the results of Ref.~\cite{uvarov2024mitigating}, yet with fewer circuit permutations required. The mitigated energy falls under the second gap (a more relevant energy scale, as the first gap separates states, which become degenerate in the thermodynamic limit) all the way to  $2p \sum\gamma_i\lesssim 5$. Interestingly, while the effect of the error mitigation is seen even beyond this point, this behavior has been observed to be problem dependent. For comparison, the figure also demonstrates the performance of LP-ZNE with 24 random permutation used and $\gamma_i^0$ sampled from the same distribution, which demonstrates inferior behavior when compared to CLP-ZNE. The inset in the figure also clearly demonstrates the bias of the LP-ZNE estimator. Overall, we conclude that the CLP-ZNE protocol can successfully mitigate even moderately large non-unital noise. 

\section{\label{sec:conclusion}Conclusion and discussion}

In the absence of error correction, algorithmic or data post-processing error mitigation methods can be used to assist the computation. A recent work \cite{uvarov2024mitigating} demonstrated that the inhomogeneity of errors in quantum computers can be exploited as a resource for error mitigation. Still, the proposed method was mostly empirical, lacking theoretical backing except in the extreme case, which required the execution of $n!$ circuits. Another limitation of the proposed method was a single-parameter noise model that could not capture the behavior of real world devices.  Thus, this method required further development. 

In this work, we proposed a Cyclic Layout Permutation-based Zero-Noise Extrapolation (CLP-ZNE) protocol that utilizes the concept of error inhomogeneity and removes most of the previous protocol limitations. The idea is based on the extrapolation of the mean expectation values, averaged across different circuit layouts, with respect to the sum of circuit errors.  
We proved that CLP-ZNE cancels all first-order noise contributions and has a quadratic bias $O(q^2)$. If the gates are affected by $d$ noise channels, the protocol requires $n(d+1)$ layouts for one-dimensional connectivity and at most $n\left(\left\lfloor\frac{n}{2}\right\rfloor d+1\right)$ layouts for arbitrary connectivity.
This provides a huge complexity reduction in terms of the required number of circuits from the previously established $n!$. 

The analytical findings were further supported by numerical simulations of the CLP-ZNE protocol for practically relevant hardware specifications of IBM Heron processor. In these results we achieved an error reduction by a factor of $8$ by considering only $2n$ layout rotations. The extrapolation in this case was simply conducted over the total circuit infidelity, readily accessible for many NISQ devices. Performing the multilinear extrapolation further improved the accuracy of the mitigated values. The applicability of the protocol was also exemplified by demonstrating the ability to mitigate circuit errors for total noise strength $\sim 1$ and beyond.

CLP-ZNE is not unique in its ability to provably mitigate the effect of gate errors up to quadratic noise terms. Such guarantees can also be established, for example, for unitary folding based ZNE. In this approach, a noisy gate $U$ is ``folded" multiple times, $U\to U(U^\dagger U)^k$, effectively increasing noise in the circuit while preserving the target unitary in the noiseless case. The most straightforward guarantee for such a protocol can be obtained under an assumption of global depolarization noise. As this channel commutes with any unitary operation, gate folding multiplicatively increases noise strength, which allows for a perfect extrapolation \cite{czarnik2021error,giurgica2020digital} using exponential fit. 
This result can be extended to more general error models that only require $U$ and $U^\dagger$ to undergo identical noise processes, yet giving rise to a polynomial bias of the extrapolation method.

Our benchmarks show that CLP-ZNE can achieve performance comparable to or better than unitary-folding ZNE. The latter methods perform best when depolarizing noise dominates, whereas CLP-ZNE becomes superior for multi-channel noise models.
This behavior could stem from greater sensitivity of unitary folding to the high-order error terms. Indeed, quadratic bias is proportional to the number of gate pairs in the circuits, which grows swiftly with the number of folds. 

Unitary folding also induces circuit runtime overheads. For example, performing $3$ circuit folds increases circuit execution time by a factor of $7$. Thus, apart from amplifying gate noise strength, this may strongly increase the role of decoherence processes, if not take the circuit out of the coherence window of a device. On the contrary, CLP-ZNE does not introduce circuit runtime overheads and only requires running the same circuit in multiple layouts. Although CLP-ZNE requires running a greater number of circuit layouts, this does not translate into a proportional increase in total sampling overhead. Indeed, CLP-ZNE performs the extrapolation over the points, each of which is obtained as an average over $n$ circuit layouts. This allows to proportionally reduce the number of shots per individual circuit to achieve the target accuracy. As established in Appendix~\ref{sec:shot noise}, the sampling overhead of CLP-ZNE remains moderate: it requires only about $10$ times more shots for the two-cycle extrapolation used in our numerical simulations to match the variance of unmitigated observable. Crucially, this factor does not scale with the number of qubits and remains constant in the small noise limit.

From a different perspective, standard ZNE techniques may be easier to implement, as they require less knowledge about the noise in the system and practically work as black-box methods. CLP-ZNE, on the other hand, can be viewed as a ``gray-box" method: it requires no knowledge about the exact noise model, but needs access to error strength characterizations. This makes this method vulnerable to  noise parameter drift. 
Nevertheless, CLP-ZNE illustrates how hardware-specific information can be incorporated into error mitigation without requiring full noise characterization. As device calibration and characterization continue to improve, such ``gray-box" approaches may provide a practical middle ground between hardware-agnostic mitigation techniques and fully noise-aware methods.

\section{Data and Code availability}
The data and the source code to reproduce all numerical simulations are available at \cite{github}.

\appendix

\section{\label{sec:appendix_parameters_vs_fidelity}Connection between noise operators and quantum channels}

The noise model described by Eqs.~\eqref{channel_general_phi} and \eqref{channel_general} encompasses different interpretations. First, it covers the case of a linear combination of different quantum channels, covering standard error models, such as depolarization, dephasing and general Pauli noise. For example, the depolarizing channel on $n$ qubits takes the form
\begin{equation}
    \mathcal{N}_p(\rho) = (1-q_p)\rho + \frac{q_p}{2^n}\Phi_{\mathds{1}}[\rho]
\end{equation}
where $\Phi_{\mathds{1}}[\rho]=\mathds{1}$ and $q_p \in \left[0,1+ \frac{1}{2^{2n}-1}\right]$ is the depolarizing parameter. 
Dephasing and arbitrary Pauli channels can be expressed similarly. 

Second, Eq.~\eqref{channel_general} can be considered as 
a linear approximation to a general quantum channel, parametrized by $d$ gate dependent parameters, $\mathcal{N}_p = \Phi(q^1_p, \dots,q^d_p)$. Indeed, in the case of weak noise
\begin{equation}
    \mathcal{N}_p = \Phi(q^1_p, \dots,q^d_p)\approx\mathcal{I}+\sum_{i=1}^dq^i_p\mathcal{E}_i,
\end{equation}
with the noise operators now given by $\mathcal{E}_i=\Phi'_{q^i}(0,\dots, 0)$. This formalism expands the applicability of Eq.~\eqref{channel_general}. For example, it allows to describe the cross-talk errors of the form $U=e^{-i\zeta tZ\otimes Z}$, which acts on an arbitrary state $\rho$ as $e^{-i\zeta t}\rho e^{i\zeta t}\approx\rho-i\zeta t[Z\otimes Z,\rho]$ for $\zeta t\ll1$. Thus, in this case, $\mathcal{E}_i(*)=-i[Z\otimes Z,*]$ and $q_i = \zeta t$. 
Similarly, small coherent under- and over-rotations $U=e^{-i q H}$, $q\ll1$ could be described by introducing  $\mathcal{E}_i(*)=-i[H,*]$.


In a specific case when the channel $\mathcal{N}$ is given as 
\begin{equation}
    \mathcal{N}=\Phi(q^1, \dots,q^d)=\Phi_1(q^1) \circ \dots\circ \Phi_d(q^d), 
\end{equation}
one finds $\mathcal{E}_i=\Phi'_{q^i}(0,\dots, 0)=\Phi'_{i}(q^i=0)$. Moreover, the associated error rates admit simple physical interpretation: due to fidelity linearity, $F_{\operatorname{avg}}(\Phi_i)\approx1+q^i F_{\operatorname{avg}}(\Phi'_{i}(0))$, implying the proportionality between the error rate and gate infidelity,
\begin{equation}
\label{eq:linear_fidelity}
    q^i \approx -\frac{1-F_{\operatorname{avg}}(\Phi_i)}{F_{\operatorname{avg}}(\Phi_i'(0)) }\propto1-F_{\operatorname{avg}}(\Phi_i).
\end{equation}
Note, however, that this observation holds only for incoherent error, when infidelity is proportional to the parameter $q$. In the case of coherent errors, the linear contribution to infidelity vanishes, implying that infidelity would grow as a second power of $q$, so Eq.~\eqref{eq:linear_fidelity} should be modified accordingly, leading to $q^i\propto \sqrt{1-F_{\operatorname{avg}}(\Phi_i)}$.

\section{\label{sec: proofs}Proof of Theorem 2}
\begin{figure}[tbh]
    \includegraphics[width=0.5\textwidth]{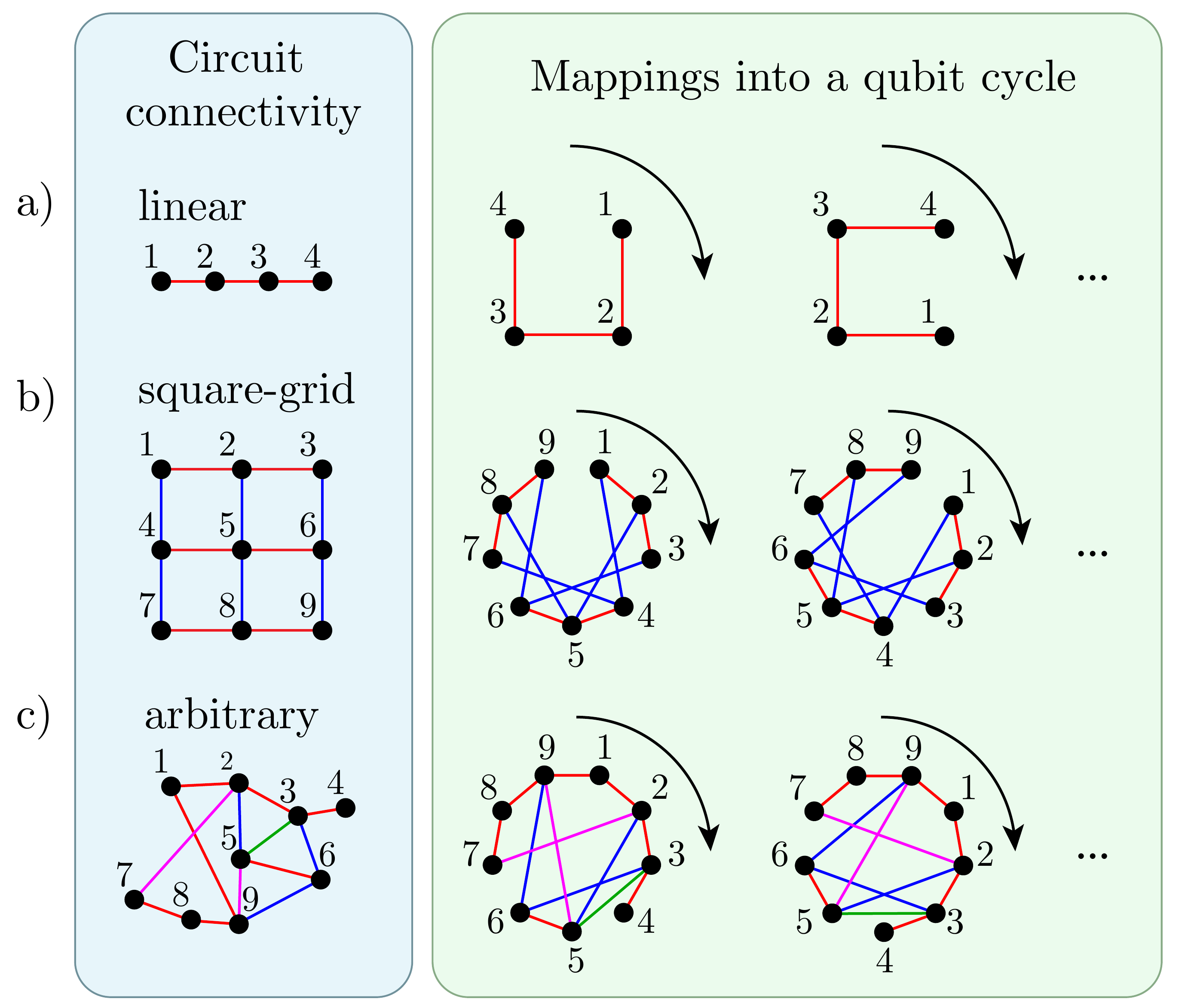}
            \caption{Cyclic layout permutations for quantum circuits with (a) linear, (b) square-grid, and (c) arbitrary connectivity. The left column displays the abstract circuit connectivity, where numbered vertices represent abstract qubits and edges denote required two-qubit interactions. The center and right panels show mappings onto a fixed cycle of $n$ physical qubits. Cyclic permutations rotate the assignment of abstract qubits to these physical sites (as indicated by curved arrows), meanwhile physical qubits remain stationary. Interactions are color-coded by gate type (graph distance along the cycle). For square-grid (b), gates are partitioned into nearest-neighbor (red) and distance-3 gates (blue). For arbitrary connectivity (c), up to $\lfloor n/2 \rfloor$ gate types appear, each in a distinct color.}
            \label{fig:explain_arbitrary_circuits}
\end{figure}

\textbf{Theorem 2 (Cyclic layout permutation-based zero-noise extrapolation for general circuit connectivity)} \\
\textit{Consider an $n$-qubit quantum circuit implemented on a device with an all-to-all connectivity. There exist a set of $M=Kd+1$ layouts $\{l_j\}_{j=1}^{M}$ mapping the circuit onto corresponding qubit cycles $\{C_j\}_{j=1}^{M}$ and a partitioning of the circuit gates into $K$ gate types $T = \bigsqcup_{k=1}^K T^k$, with
\begin{itemize}
    \item $K=2$ for square-grid circuit connectivity;
    \item $K \le  \lfloor \frac{n}{2} \rfloor$ for arbitrary circuit connectivity,
\end{itemize}
which allow executing zero-noise extrapolation. Define $\langle E\rangle_{\mathcal{C}(l_j)}$ as the expectation values averaged over the sets of cyclic layout permutations $\mathcal{C}(l_j)$. 
Then a linear extrapolation of  $\langle E\rangle_{\mathcal{C}(l_j)}$ to the level of zero noise with respect to the mean total error rates $\left\langle \sum_{g\in T^k}q_g^i\right\rangle_{\mathcal{C}(l_j)}$, provides an estimate $E_{\operatorname{mit}}$ that approximates the noiseless expectation value $E_0$  up to the terms quadratic in $q$, }
\begin{equation}
    E_{\operatorname{mit}}=E_0+O(q^2).
\end{equation}
\begin{proof}
We begin by partitioning circuit gates into $K$ distinct types and then construct the required circuit layouts.

\noindent
\textbf{Gate partitioning.}
For the square-grid case, number the abstract circuit qubits row-wise, starting from the top row, as depicted in Fig.~\hyperref[fig:explain_arbitrary_circuits]{7(b)} (left panel). For arbitrary connectivity, the abstract qubits are assigned an arbitrary fixed ordering, as exemplified in Fig.~\hyperref[fig:explain_arbitrary_circuits]{7(c)} (left panel). 

Partition the gates into $K$ equivalence classes (gate types) $T = \bigsqcup_{k=1}^K T^k$ based on the cyclic distance $d_{ij}=\min\{|i-j|, n-|i-j|\}$ between qubits the gate acts on. Two gates belong to the same class if and only if their respective qubit distances are equal. For the $w \times h$ square grid, only two classes appear: nearest-neighbor and distance-$w$ connections, yielding $K=2$ (Fig.~\hyperref[fig:explain_arbitrary_circuits]{7(b)}). For arbitrary connectivity, up to $K \leq \lfloor n/2 \rfloor$ classes may be present (Fig.~\hyperref[fig:explain_arbitrary_circuits]{7(c)})  with exact equality holding if all distances are utilized.

\noindent
\textbf{Layout construction.} 
Select $M=Kd+1$ distinct qubit cycles $\{C_j\}_{j=1}^{M}$ on the device, each containing $n$ qubits. For each cycle $C_j$, the layout $l_j$ is defined by mapping the ordered abstract qubits clockwise (or counter-clockwise) onto the physical qubits of $C_j$, starting from an arbitrary physical site (Fig.~\ref{fig:explain_arbitrary_circuits}, center panel). From each $l_j$, we generate the set $\mathcal{C}(l_j)$ of $n$ cyclically permuted layouts by uniformly rotating the abstract-to-physical assignment along $C_j$ (Fig.~\ref{fig:explain_arbitrary_circuits}, center and right panels).

\noindent
\textbf{Layout averaging.} By construction, under a layout $l_j$, all gates in $T^k$ will be executed on physical qubits separated by a fixed graph distance along the cycle $C_j$. Consequently, each subset $T^k$ corresponds to a specific chord type; chords of the same type connect qubits equidistant along the cycle and can be mapped onto one another via cyclic rotations.
Upon rotations of this layout over the cycle, every abstract gate $g \in T^k$ is executed exactly once on each of the $n$ physical chords of type $k$ in cycle $C_j$. This makes the mean noise strengths $\langle q_g^i \rangle_{\mathcal{C}(l_j)}$ independent of the specific gate $g \in T^k$.

With this observation, the noisy expectation values from Eq.~\eqref{energy_with_layou_avg} become
\begin{equation}
    \langle E \rangle_{\mathcal{C}(l_j)}= E_0+\sum_{i=1}^d\sum_{k}\Delta_{i,k}\left\langle  \sum_{g\in T^k}q_g^i\right\rangle_{\mathcal{C}(l_j)}+O(q^2),
    \label{eq:lemma1_output_general}
\end{equation}
where $\Delta_{i,k}=\left(\sum_{g\in T^k} E_g^i\right)/|T^k|$ are the average perturbations originating from the error sources $\mathcal{E}_i$ for the subset of gates $T^k$. 

The right-hand side of Eq.~\eqref{eq:lemma1_output_general} contains $dK$ unknown values $\Delta_{i,k}$ together with the unknown $E_0$, giving $dK+1$ unknowns in total. Thus, Eq.~\eqref{eq:lemma1_output_general} enables linear ZNE over the features $\left\langle \sum_{g\in T^k} q_g^i \right\rangle_{\mathcal{C}(l_j)}$ when $M = dK+1$ sets of cyclic permutations are used.

\noindent
\textbf{Linear regression.} To perform the linear regression, we introduce a design matrix $\mathbf{X}$ \cite{Timm2002} with elements
\begin{equation}
    \mathbf{X}_{j\alpha} = 
    \begin{cases}
        1, & \alpha = (0,0), \\
        \left\langle \sum_{g\in T^k} q_g^i \right\rangle_{\mathcal{C}(l_j)}, & \alpha \ne (0,0). 
    \end{cases}
    \label{eq:design_matrix_arb}
\end{equation}
The composite index $\alpha = (i,k)$ is used here  to enumerate all pairs $(i,k)$ for $1 \leq i \leq d, \; k \in \{1,\ldots,K\}$. Equation~\eqref{eq:lemma1_output_general} can then be rewritten in matrix form as:
\begin{equation}
    \langle E \rangle_{\mathcal{C}(l_j)} = \sum_{\alpha=0}^{dK} \mathbf{X}_{j\alpha} \Delta_\alpha + O(q^2),
\end{equation}
where $\Delta_0 = E_0$ and $\Delta_{(i,k)} = \Delta_{i,k}$ for $\alpha \geq 1$. In vector notation, $\vec{E} = \mathbf{X} \vec{\Delta} + O(q^2)$, where $\vec{E}$ is the $M$-dimensional vector of expectation values and $\vec{\Delta} = (E_0, \Delta_1, \ldots, \Delta_{dK})^T$. 
This dependence can be fit with a linear model which, after minimizing the square distance to the noisy expectation values, yields the mitigated expectation given by the first component of
$\vec{\Delta} = \mathbf{X}^{-1} \vec{E}$, namely
\begin{equation}
    E_{\operatorname{mit}} = \sum_{j=1}^M (\mathbf{X}^{-1})_{0j} \langle E \rangle_{\mathcal{C}(l_j)}.
\end{equation}
    
Finally, we note that the first row of $\mathbf{X}^{-1}$ remains finite for $q\to0$, implying that $E_{\operatorname{mit}} = E_0+\sum_{j=1}^M (\mathbf{X}^{-1})_{0j} O(q^2) = E_0+O(q^2)$, which finishes the proof. 
\end{proof}
Similar to the case discussed in the main text, if necessary, the matrix $\mathbf{X}$ can be appended with extra rows to ensure full column rank. Then the extrapolation would again retrieve $E_0$ using the Moore–Penrose pseudoinverse instead of $\mathbf{X}^{-1}$.

\section{Bias analysis}
\label{bias_estimate}
In this appendix, we provide a bound on the leading-order bias of the CLP-ZNE protocol. Although Theorems 1 and 2 guarantee that the mitigated estimate deviates from the noiseless value only by terms quadratic in the noise strength, the following analysis goes further by explicitly bounding the magnitude of these residual quadratic contributions. For simplicity, we focus on bounding the quadratic terms for Theorem 1; the same argument extends to Theorem 2 with only minor modifications.

Taking the trace of the Eq.~\eqref{scary_full_noise_state} with an observable $H$, we find that the noisy expectation value for a specific layout $l$ is given by $E_l = E_0 + \sum_{s \neq 0} a_{l,s} E_s$, where $a_{l,s} = \prod_{i=1}^d \prod_{g \in T} (q_{l(g)}^i)^{s_g^i}$ and $E_s = \operatorname{Tr}(\rho_s H)$. The CLP-ZNE mitigated estimator is defined as $E_{\text{mit}} = \sum_{j=1}^{d+1} w_j \langle E \rangle_{\mathcal{C}(l_j)}$, with weights $w_j = (X^{-1})_{0j}$ and $X$ defined in Eq.~\eqref{eq:design_matrix}. By construction, the design matrix $X$ is chosen so that all first-order contributions to the bias vanish. Defining $R_l^{(2)}=\sum_{|s| =2} a_{l,s} E_s$, the second-order contribution to the bias is given by
\begin{equation}
    \text{Bias}^{(2)} = \sum_{j=1}^{d+1} w_j \langle R^{(2)} \rangle_{\mathcal{C}(l_j)},
    \label{eq:bias_exact}
\end{equation}
where $\langle R^{(2)} \rangle_{\mathcal{C}(l_j)}$ is the second-order remainder averaged over the cyclic permutations of layout $l_j$, and $|s|$ denotes the number of non-zero elements of $s$.

Let $Q_l = \sum_{i=1}^d \sum_{g \in T} q_{l(g)}^i$ be the total first-order error rate for layout $l$. Crucially, the physical noise model enforces the constraint $\sum_{i=1}^d s_g^i \le 1$ for all $g \in T$, meaning at most one noise channel can act per gate. For the second-order terms, this constraint implies that the two noise operators must act on distinct gates. Consequently, the sum of the second-order coefficients $a_{l,s}$ is bounded by the cross terms in the square of the total error rate, yielding
\begin{equation}
    \sum_{|s|=2} a_{l,s} \le \frac{1}{2} \left( \sum_{i=1}^d \sum_{g \in T} q_{l(g)}^i \right)^2 = \frac{1}{2} Q_l^2.
    \label{eq:prob_bound}
\end{equation}

Let $\Delta E_{\max} = \max_{|s|=2} |E_s|$ be the maximum second-order energy perturbation. By the triangle inequality, the absolute value of the second-order remainder for a specific layout is bounded by
\begin{equation}
    \big| R_l^{(2)} \big| \le \left( \sum_{|s|=2} a_{l,s} \right) \Delta E_{\max} \le \frac{1}{2} Q_l^2 \Delta E_{\max}.
    \label{eq:remainder_bound}
\end{equation}

To express this in terms of global circuit parameters, we bound $Q_l \le d \|T\| q_{\max}$, where $q_{\max} = \max_{i, g, l} q_{l(g)}^i$. Substituting this into Eq.~\eqref{eq:bias_exact} yields the bound on the second-order systematic bias
\begin{equation}
    |\text{Bias}^{(2)}| \le \frac{1}{2} \Delta E_{\max} (d \|T\| q_{\max})^2 \sum_{j=1}^{d+1} \big| (X^{-1})_{0j} \big|.
    \label{eq:bias_final}
\end{equation}

Since the extrapolation weights $(X^{-1})_{0j}$ remain finite as $q \to 0$, the second-order bias remains well bounded and negligible in the low-noise regime, validating the efficiency of the CLP-ZNE protocol.

\section{\label{sec:shot noise}Shot noise  and sampling overhead}
\subsection{Shot-noise analysis}
In practice, estimating expectation values on a quantum computer requires repeated measurements (or ``shots''). Even in the absence of hardware noise, a finite number of shots introduces statistical uncertainty in the estimated expectation value, a phenomenon commonly referred to as shot noise. The empirical estimates of $\langle E\rangle_j \equiv \langle E\rangle_{\mathcal{C}(l_j)}$ used as inputs to the extrapolation procedure are therefore subject to this statistical error, which inevitably degrades the accuracy of the mitigated result $E_{\text{mit}}$.

To quantify this effect, we consider a regime where shot noise dominates the total error in $E_{\text{mit}}$, and systematic errors arising from first-order perturbation approximation of noisy expectation values are negligible. Under this assumption, the measured expectation values can be modeled as
\begin{equation}
    \langle \hat{E}\rangle_j = E_0 + \sum_{i=1}^d \Delta_i \left\langle \sum_{g\in T} q_g^i \right\rangle_{\mathcal{C}(l_j)} + \varepsilon_j,
    \label{eq:finite_stat_model}
\end{equation}
where $\varepsilon_j \sim \mathcal{N}\bigl(0,\,\mathrm{Var}(\langle \hat{E}\rangle_j)\bigr)$ represents the shot-noise-induced error for the $j$th cycle, and $\langle \hat{E}\rangle_j$ denotes the empirical estimate of $\langle E\rangle_j$ obtained from finite sampling.

Performing a least-squares fit of the model in the Eq.~\eqref{eq:finite_stat_model} yields a mitigated expectation value $E_{\text{mit}}$ whose variance is given by

\begin{equation}
    \mathrm{Var}(E_{\text{mit}}) = \bigl[(X^\top X)^{-1} X^\top \Sigma X (X^\top X)^{-1}\bigr]_{00},
    \label{eq:variance_propagation}
\end{equation}
where $X$ is the design matrix (defined in Eq.~\eqref{eq:design_matrix}), and $\Sigma = \mathrm{diag}\bigl(\mathrm{Var}(\langle \hat{E}\rangle_1), \dots, \mathrm{Var}(\langle \hat{E}\rangle_{d+1})\bigr)$.

To evaluate the shot-noise variance for a given circuit layout $l$, we decompose the Hamiltonian into $A$ groups of mutually commuting Pauli terms
\begin{equation}
    H = \sum_{\alpha=1}^A H_\alpha,
\end{equation}
where each $H_\alpha$ contains only commuting Pauli strings and can be measured simultaneously. Let $N_\alpha$ be the number of measurement shots for group $\alpha$. The variance of the energy estimator for circuit layout $l$ is
\begin{equation}
    \mathrm{Var}(\hat{E}_l) = \sum_{\alpha=1}^A \frac{\mathrm{Var}(H_\alpha)_l}{N_\alpha}
    = \sum_{\alpha=1}^A \frac{\langle H_\alpha^2 \rangle_l - \langle H_\alpha \rangle_l^2}{N_\alpha},
    \label{eq:variance_per_layout}
\end{equation}
where $\langle \cdot \rangle_l$ denotes the expectation value with respect to the quantum state prepared by the circuit with layout $l$. Let $N_{\text{mit}}$ denote the total number of measurement shots allocated across the entire CLP-ZNE protocol. Assuming uniform shot allocation across all layouts and commuting groups, the number of shots per group per layout is $N_\alpha = N_{\text{mit}} / ((d+1)mA)$. Substituting this allocation into Eq.~\eqref{eq:variance_per_layout}, the variance becomes
\begin{equation}
    \mathrm{Var}(\hat{E}_l) = \frac{m A (d+1)}{N_{\text{mit}}} \sum_{\alpha=1}^A \bigl( \langle H_\alpha^2 \rangle_l - \langle H_\alpha \rangle_l^2 \bigr).
    \label{eq:variance_eq}
\end{equation}

 The shot-noise variance for each empirical estimate $\langle \hat{E}\rangle_j$ arises from averaging over $m$ independent sets of measurements, where $m = |\mathcal{C}(l_j)|$ (with $|\cdot|$ denoting set cardinality). Specifically,
\begin{equation}
    \mathrm{Var}(\langle \hat{E}\rangle_j) = \frac{1}{m^2} \sum_{l \in \mathcal{C}(l_j)} \mathrm{Var}(\hat{E}_l).
    \label{eq:sampling_variance}
\end{equation}
This layout-dependent expression clarifies that $\mathrm{Var}(\langle \hat{E}\rangle_j)$ may vary due to differences in the underlying quantum states; though, when circuit noise is weak, these variations are typically mild, thus we denote $\mathrm{Var}(\hat{E}_l) \approx \mathrm{Var}(\varepsilon)$ for all $l$. This leads to the homoscedastic approximation

\begin{equation}
    \mathrm{Var}(E_{\text{mit}}) = \frac{\mathrm{Var}(\varepsilon)}{m}\, \bigl[(X^\top X)^{-1}\bigr]_{00}.
    \label{eq:homoscedastic_1}
\end{equation}

Combining Eqs.~\eqref{eq:homoscedastic_1} and \eqref{eq:variance_eq}, within homoscedastic approximation we obtain 
\begin{equation}
    \mathrm{Var}(E_{\text{mit}})  
    = \frac{A (d+1) \big[(X^\top X)^{-1}\big]_{00} }{N_{\text{mit}}} \sum_{\alpha=1}^A \bigl( \langle H_\alpha^2 \rangle - \langle H_\alpha \rangle^2 \bigr) .
    \label{eq:total_error_bound}
\end{equation}
For single error source extrapolation ($d=1$), this simplifies to
\begin{equation}
    \mathrm{Var}(E_{\text{mit}}) = \frac{A (d+1) \ (e_1^2 + e_2^2)}{N_{\text{mit}}\, (e_1 - e_2)^2} \sum_{\alpha=1}^A \bigl( \langle H_\alpha^2 \rangle - \langle H_\alpha \rangle^2 \bigr) ,
    \label{eq:single_channel_error}
\end{equation}
where $e_1 = \left\langle \sum_{g\in T} q_g^1\right\rangle_{\mathcal{C}{(l_1)}}$ and $e_2 = \left\langle \sum_{g\in T} q_g^1\right\rangle_{\mathcal{C}{(l_2)}}$.
These expressions show that the shot noise error in the mitigated result scales as $\sigma_{E_{\text{mit}}} =\sqrt{\mathrm{Var}(E_{\text{mit}})}\propto1/\sqrt{N_{\text{mit}}}$, consistent with known results from statistics. Consequently, to achieve a target error tolerance $\epsilon$, the total number of measurements must scale as $O(1/\epsilon^2)$.

\subsection{Sampling overhead}
This result allows to estimate the sampling overhead required to maintain a target variance. It is defined as the ratio between the total number of measurements for the error mitigation protocol ($N_{\text{mit}}$) and noisy observable estimation ($N_{\text{noisy}}$) required to achieve equal variance, i.e., $\operatorname{Var}(E_{\text{mit}}) = \operatorname{Var}(E_{\text{noisy}})$. The latter variance is derived analogously to Eq.~\eqref{eq:variance_per_layout}, assuming a total of $N_{\text{noisy}}$ shots allocated uniformly across the $A$ commuting groups

\begin{equation}
    \operatorname{Var}(E_{\text{noisy}})
    = \frac{A}{N_{\text{noisy}}}
      \sum_{\alpha=1}^{A}\bigl( \langle H_{\alpha}^{2} \rangle - \langle H_{\alpha} \rangle^{2} \bigr).
    \label{eq:var_noisy}
\end{equation}
Thus, the sampling overhead of CLP-ZNE reads
\begin{equation}
    C_{\text{CLP}} = \frac{N_{\text{mit}}}{N_{\text{noisy}}}
                  = (d+1)\,\bigl[(X^{\top}X)^{-1}\bigr]_{00}.
    \label{eq:sampling_overhead}
\end{equation}
For the practically important case of a single dominant error source ($d=1$) and two extrapolation points Eq.~\eqref{eq:sampling_overhead} simplifies to
\begin{equation}
    C_{\text{CLP}} = \frac{2(e_{1}^{2}+e_{2}^{2})}{(e_{2}-e_{1})^{2}},
    \label{eq:overhead_two_cycles}
\end{equation}
where $e_{j} = \bigl\langle \sum_{g\in T} q_{g} \bigr\rangle_{\mathcal{C}(l_{j})}$ are the mean total error rates for the two cycles. In the numerical simulations presented in this work, the two cycles are chosen such that $e_{2}\approx 2e_{1}$. Substituting this ratio into Eq.~\eqref{eq:overhead_two_cycles} yields $C_{\text{CLP}}\approx 10$, indicating a moderate sampling overhead compared to the unmitigated case.

\end{document}